\documentclass[twocolumn]{aa} 

\usepackage{txfonts}
\usepackage{graphicx}
\def\xmm{{\it XMM-Newton}}
\def\intgr{{\it INTEGRAL}}

\begin{document}

   \title{GMRT observations of the Ophiuchus galaxy cluster}

 \author{M. Murgia\inst{1}
          \and
          D. Eckert\inst{2}
          \and
          F. Govoni\inst{1}
         \and
          C. Ferrari\inst{3}
          \and
          M. Pandey-Pommier\inst{4}
          \and
          J. Nevalainen\inst{5}
          \and
          S. Paltani\inst{6}
          }
   \institute{
              INAF - Osservatorio Astronomico di Cagliari, Poggio dei Pini, Strada 54, 09012 Capoterra (CA), Italy
              \and
              INAF - IASF-Milano, Via Bassini 15, 20133 Milano, Italy
              \and              
	UNSA, CNRS UMR 6202 Cassiop\'ee, Observatoire de la C\^ote d'Azur, Nice, France
        \and
 CRAL - Observatoire de Lyon, 9, Avenue Charles Andr\'e, 69230 Saint-Genis-Laval, France
\and
Department of Physics, p.o. box 48, 00014 University of Helsinki, Finland
\and
ISDC, Universit\'e de Gen\`eve, 16, ch. d'Ecogia, 1290 Versoix Switzerland
}

   \date{Received; accepted}

% \abstract{}{}{}{}{} 
% 5 {} token are mandatory
 
  \abstract
  {} 
  {Very Large Array observations at 1477 MHz revealed the presence of a
    radio mini-halo surrounding the faint central point-like radio
    source in the Ophiuchus cluster of galaxies. In this work we
    present a study of the radio emission from this cluster of
    galaxies at lower radio frequencies.}
  {We observed the Ophiuchus cluster at 153, 240, and 614 MHz with the
    Giant Metrewave Radio Telescope.}
  {The mini-halo is clearly detected at 153 and 240 MHz, the frequencies at
    which we reached the best sensitivity to the low-surface brightness
    diffuse emission, while it is not detected at 610 MHz because of the too low signal-to-noise ratio at this frequency. 
The most prominent feature at low frequencies is a patch of 
diffuse steep spectrum emission located at about 5\arcmin~south-east from the cluster centre.
By combining these images with that at 1477 MHz, we
    derived the spectral index of the mini-halo. Globally, the mini-halo has a low-frequency spectral index of $\alpha_{240}^{153}\simeq 1.4\pm0.3$ and an high-frequency spectral index of $\alpha_{1477}^{240}\simeq 1.60\pm 0.05$. Moreover, we measure a systematic increase of the high-frequency spectral index 
with radius: the azimuthal radial average of  $\alpha_{1477}^{240}$ increases from about 1.3, at the cluster centre, up to about 2.0 in the mini-halo outskirts.}
  {The observed radio spectral index is in agreement with that obtained by 
   modeling the non-thermal hard X-ray emission in this
    cluster of galaxies. We assume that the X-ray component arises from inverse-Compton 
scattering between the photons of the cosmic microwave background and a population of non-thermal electrons 
which are isotropically distributed and whose energy spectrum is a power law with index $p$. We derive that
the electrons energy spectrum should extend from a minimum Lorentz factor of $\gamma_{min}\lesssim 700$ up to 
a maximum Lorentz factor of $\gamma_{max}\simeq 3.8 \times 10^{4}$ with an index $p=3.8\pm 0.4$. The 
volume-averaged strength for a completely disordered intra-cluster magnetic field is $B_V\simeq 0.3\pm 0.1\, \mu$G.}


\titlerunning{GMRT observations of the Ophiuchus galaxy cluster}

   \maketitle

\section{Introduction}

Galaxy clusters, the largest gravitationally bound structures in the 
Universe, are still forming at present epoch by merging of nearly 
equal-mass systems or accretion of groups and field galaxies. They are 
excellent laboratories to study the baryonic cosmic fraction, as well as 
the interplay between baryonic and dark matter in the formation and 
evolution process of large scale structures (e.g. Arnaud et al. 2009; 
Kravtsov et al. 2009). 
In the last twenty years important progresses have been made in the study 
of cluster of galaxies, of the thermal intra-cluster medium (ICM) and of 
their interaction (e.g. Boselli \& Gavazzi 2006; Markevitch \& Vikhlinin 
2007). Much less is instead known about the physical properties and the 
origin of a non-thermal intra-cluster component (relativistic electrons 
with energies of $\simeq 10$ GeV spiralling in magnetic fields of few 
$\mu$Gauss) that has been discovered and studied mostly through deep 
radio observations (see e.g. Ferrari et al. 2008 and references therein). 
However, it is now clear that the impact of the non-thermal component 
in the physics and thermo-dynamical evolution of galaxy clusters cannot 
be neglected anymore (e.g. Dursi \& Pfrommer 2008; Parrish et al. 2009).

Intra-cluster relativistic electrons radiate through synchrotron 
emission in the radio domain, but also through inverse Compton 
scattering of  cosmic microwave background (CMB) photons in the hard X-ray 
(HXR) band. The diffuse non-thermal component is now well detected at radio 
wavelengths in about 30 clusters (Giovannini et al. 2009). 
Only a few X-ray satellites allowed possible but controversial detection 
of a hard tail in the X-ray spectrum of about 10 clusters 
(see e.g. Fusco-Femiano et al. 2003; Nevalainen et al. 2004; Rephaeli et al. 2008). Very recent 
results are either in agreement with a HXR non-thermal detection (e.g. 
Eckert et al. 2008) or suggest a possible thermal origin of the detected 
HXR emission (e.g. Kawano et al. 2009).

\begin{table*}[t]
\caption{Details of the GMRT observations.}
\begin{center}
\begin{tabular}{cccccccc}
\hline
\noalign{\smallskip}
 Frequency  & Sideband    & Bandwidth/sideband & Polarization & Observation Time  & Date       & Beam     & rms             \\
    (MHz)   & (LSB,USB)  & (MHz)     &              &  (hours)     &            & (arcsec)    & (mJy/beam)        \\
\noalign{\smallskip}
\hline
\noalign{\smallskip}
153         &   BOTH, BOTH &  8        & LL/RR, LL/RR    & 5.2+4.7       & 21, 22 August 2008  &  $31.1 \times 22.6$  & 5     \\
240         &   BOTH, USB  &  8        & LL/RR, LL       & 4.6+4.3       & 23, 28 August 2008 &  $18.7 \times 15.9$  & 1.1     \\
614         &   BOTH       & 16        &   RR            & 4.3           & 28 August 2008     &  $7.0 \times 7.0 $    & 0.25  \\
\noalign{\smallskip}
\hline
\end{tabular}
\label{tab1}
\end{center}
\end{table*}

The Ophiuchus cluster ($z=0.028$, Johnston et al. 1981) is one of  
the brightest cluster of galaxies in the X-ray band. It is an 
extremely interesting target for non-thermal cluster studies, since it 
shows evidence of both radio and, possibly, HXR emission (Eckert et al. 
2008; Govoni et al. 2009; Murgia et al. 2009; Nevalainen et al. 2009).
The dynamical state of the Ophiuchus cluster has been strongly debated 
in the last years. A recent {\it Chandra} study (Million et al. 2009) shows
evidence of a recent merger event in the central region of the cluster 
of 8 $\times$ 8 arcmin$^2$ (but see also Fujita et al. 2008 for an opposite conclusion 
based on {\it Suzaku} data). In addition several clusters and groups 
of galaxies have been detected within a distance of 8$^{\circ}$ from 
the cluster centre, indicating that Ophiuchus is in a supercluster 
environment (Wakamatsu et al. 2005). 

By analyzing Very Large Array (VLA) data of Ophiuchus at 1477 MHz, Govoni et al. 
(2009) recently detected a radio mini-halo surrounding the faint central 
point-like radio source. Radio 
mini-halos are diffuse steep-spectrum ($\alpha > 1$; $S_{\nu}\propto 
\nu^{- \alpha}$) sources, permeating the central regions of relaxed, 
cool-core, galaxy clusters. They usually surround a radio galaxy. These 
diffuse radio sources are extended on a moderate scale (typically 
$\simeq$ 500 kpc) and, in common with large-scale halos observed in 
merging clusters of galaxies, have a steep spectrum and a very low 
surface brightness.  As a consequence of their relatively small angular 
size and possibly strong radio emission of the central radio galaxy, 
radio mini-halos are very elusive sources and our current observational 
knowledge on mini-halos is limited to only a handful of well-studied 
clusters.

Based on current observational and theoretical analyses, radio emission 
from mini-halos would be due to a population of relativistic electrons 
ejected by the central AGN and re-accelerated by MHD turbulence, whose 
energy is, in turn, supplied by the cluster cooling-core (Gitti et al. 
2004). Recent analysis of the most X-ray luminous cluster (RX 
J1347-1145) suggest that additional energy for electron re-acceleration 
in mini-halos might be provided by sub-cluster mergers that have not 
been able to destroy the central cluster cooling-core (Gitti et al. 
2007). Ophiuchus is the second known cluster showing a radio mini-halo, 
as well as a cool-core that has survived a possible recent merging event 
(Nevalainen et al. 2009; Million et al. 2009). Indeed, Burns et al. (2008)
simulated the formation and evolution of galaxy clusters,
and showed that cool-core clusters can accrete mass 
over time and grew slowly via hierarchical mergers; 
when late mergers occur, the cool-cores survive the collisions.

Eckert et al. (2008) measured a high confidence level (6.4 $\sigma$) HXR 
excess in Ophiuchus through {\it INTEGRAL} observations.  This emission 
may be of non-thermal origin, caused presumably by Compton scattering of 
cosmic microwave background radiation by the relativistic electrons 
responsible for the mini-halo emission (see e.g., Rephaeli et al. 2008, 
Petrosian et al. 2008, and references therein for reviews). Alternative 
explanations have also been put forward (Profumo 2008, P{\'e}rez-Torres 
et al. 2009, Colafrancesco \& Marchegiani 2009). The HXR excess 
detection in Ophiuchus was recently confirmed by Nevalainen et al. 
(2009). In addition, their joint {\it INTEGRAL} and {\it XMM} analysis 
partly reconciled the previous discrepancy between the results by Eckert 
et al. (2008) and the upper limits on HXR flux obtained by Ajello et al. 
(2009) through Swift/BAT data.

Ophiuchus is thus one of the few clusters of galaxies in which the 
non-thermal component is revealed both in the radio and in the HXR 
bands. For this reason, it is particularly interesting to investigate 
the radio spectrum of the mini-halo. By combining this information with 
the observed properties of the hard X-ray emission it would be possible 
to derive important constraints on the energy spectrum of the 
synchrotron electrons. In particular, by assuming that the synchrotron 
emission and the hard X-ray excess are co-spatial and produced by the 
same population of relativistic electrons, their comparison would allow 
the determination of the cluster magnetic field (Nevalainen et al. 
2009).

In this work we present a study of the radio emission from the Ophiuchus 
cluster of galaxies at low radio frequencies. We observed the Ophiuchus 
cluster at 153, 240, and 614 MHz with the Giant Metrewave Radio 
Telescope (GMRT).
Throughout this paper we assume a $\Lambda$CDM cosmology with $H_0$ = 71 
km s$^{-1}$Mpc$^{-1}$, $\Omega_m$ = 0.27, and $\Omega_{\Lambda}$ = 0.73. 
At the distance of Ophiuchus ($z=0.028$), 1\arcsec~ corresponds to 0.55 
kpc.

\section{GMRT observations}

We observed the cluster of galaxies Ophiuchus using the GMRT at the 
frequencies of 153, 240, and 614 MHz (program OJ1712).  The GMRT 
antennas were pointed at RA=17$^{h}$12$^{m}$28$^{s}$ and DEC=$-$23\degr 
22\arcmin 06\arcsec (J2000).  The visibilities have been acquired in 
spectral line mode in order to reduce the bandwidth smearing effect and 
to facilitate the excision of narrow band radio frequency interferences 
(RFIs).

We summarize the details of the observations in Table\,\ref{tab1}, where 
we provide the frequency and total bandwidth, observation date, total 
time on source, beamwidth (FWHM) of the full array, and rms 
level (1$\sigma$) in the full-resolution images. Calibration and imaging 
were performed with the NRAO Astronomical Image Processing System 
(AIPS).

The Ophiuchus cluster have been already observed with the GMRT
at 240 and 610 MHz by P{\'e}rez-Torres et al. (2009) who did not detect
the diffuse mini-halo emission at the noise level of their images. The observations
presented here, however, have on average an exposure time about a factor 4.5 longer
and hence they permit to go deeper in sensitivity.

\begin{figure*}[t]
\centering
\includegraphics[width=19 cm]{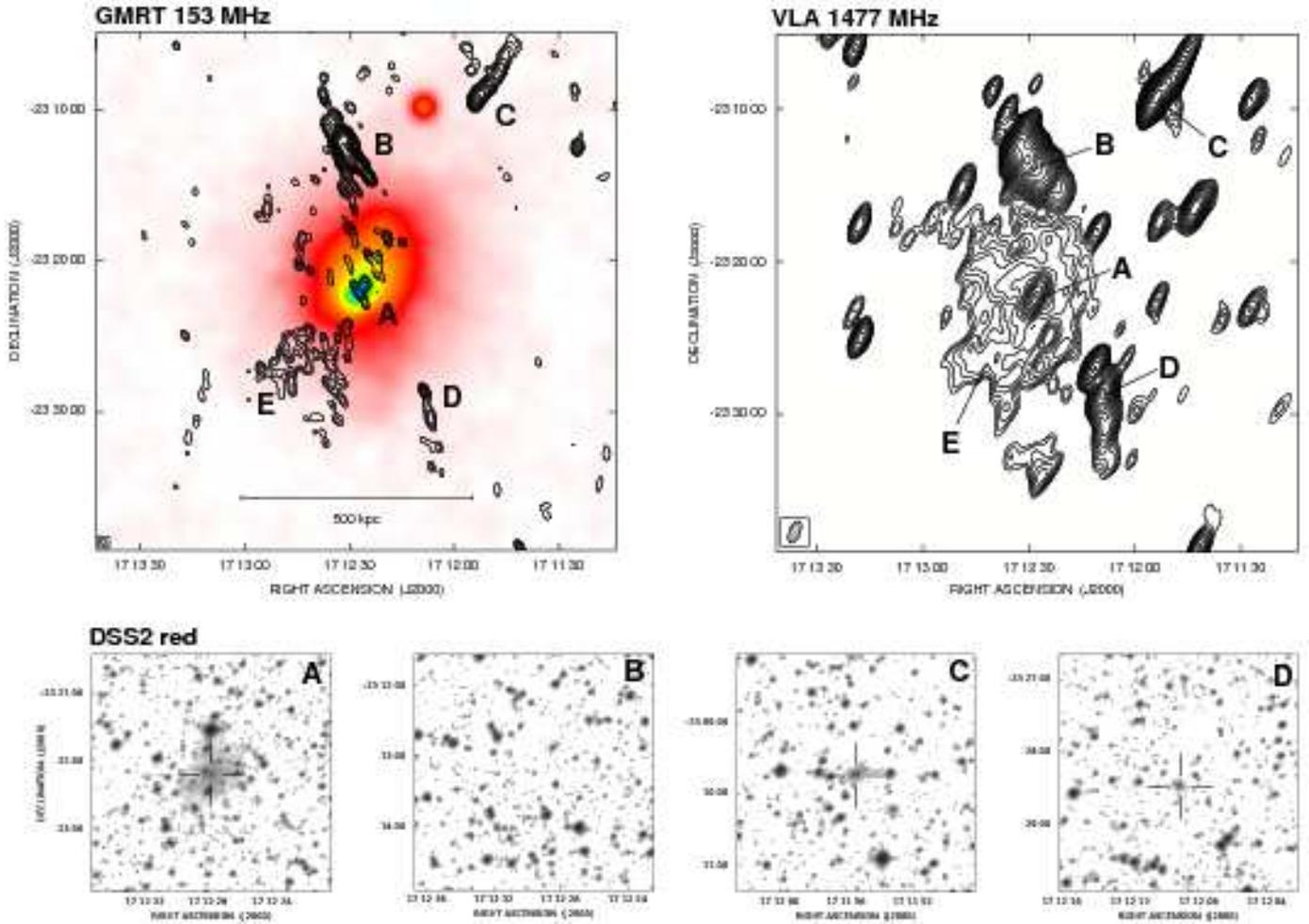}
\caption{Top left: GMRT radio iso-contours at 153 MHz at full-resolution overlaid to the ROSAT PSPC X-ray image of the Ophiuchus 
galaxy cluster. The radio image has an FWHM beam of $37.1\arcsec \times22.6\arcsec$ with PA=35.1\degr and a noise level 
of 5 mJy/beam (1$\sigma$). Contours start at 15 mJy/beam ($3\sigma$) and scale by a factor of $\sqrt{2}$. Top right: iso-contours
 at the frequency of 1477 from the VLA image by Govoni et al. (2009). The image as an FWHM beam of $91.4\arcsec \times40.3\arcsec$ with PA=-24.4\degr. Contours start at 0.3 mJy/beam ($3\sigma$) and increase by $\sqrt{2}$. Bottom panels: finding charts from DSS2$-$red for sources A, B, C, and D. No obvious optical identification exists for radio 
source B.}
\label{fig1}
\end{figure*}

\subsection{153 MHz} 

The observations were performed using central frequency of 153 MHz and a 
bandwidth of 8 MHz for both the upper and lower sideband (USB and LSB). 
The data were collected in spectral line mode with 128 spectral channels 
of 62.5 kHz in width.  The observations consist of two distinct runs of 
about 5.2 and 4.7 hours on source, performed on 21 and 22 August 2008, 
respectively.  The flux density scale and the bandpass were calibrated 
for both runs by using the primary calibrator 3C\,286. 
An initial amplitude solution for the bandpass calibrators was first 
obtained for a central channel free of RFIs.  This channel was then 
used as reference in task BPASS in order to derive a bandpass solution. 
The bandpass solution was visually inspected 
and the most obvious RFIs were carefully excised. This 
procedure was repeated several times until a refined bandpass solution 
was obtained. Task FLGIT was applied to the whole dataset and about 30\% 
of the data were automatically removed because of the contamination from 
strong RFIs. We run task SPLAT to apply the bandpass calibration and to 
reduce the number of channels from 128 to 6 channels of 1 MHz width 
each. We calibrated the final data set in both phase and amplitude. The 
phase calibration was completed by using the secondary calibrators 
1830-360 and 1833-210, observed at intervals of $\sim$30 minutes. 
Low-level residual RFIs were carefully removed from the 6-channels 
dataset by visually inspection and finally the Ophiuchus 
data were extracted with task SPLIT and imaged.  
Several cycles of 
self-calibration were applied in order to remove residual phase 
variations. 

We run task IMAGR in 3D mode over a mosaicing of  
slightly overlapping fields in order to account for the non-coplanarity 
of the incoming wavefront within the large primary beam of $\sim 
3\degr$.

\begin{figure*}[t]
\centering
\includegraphics[width=18 cm]{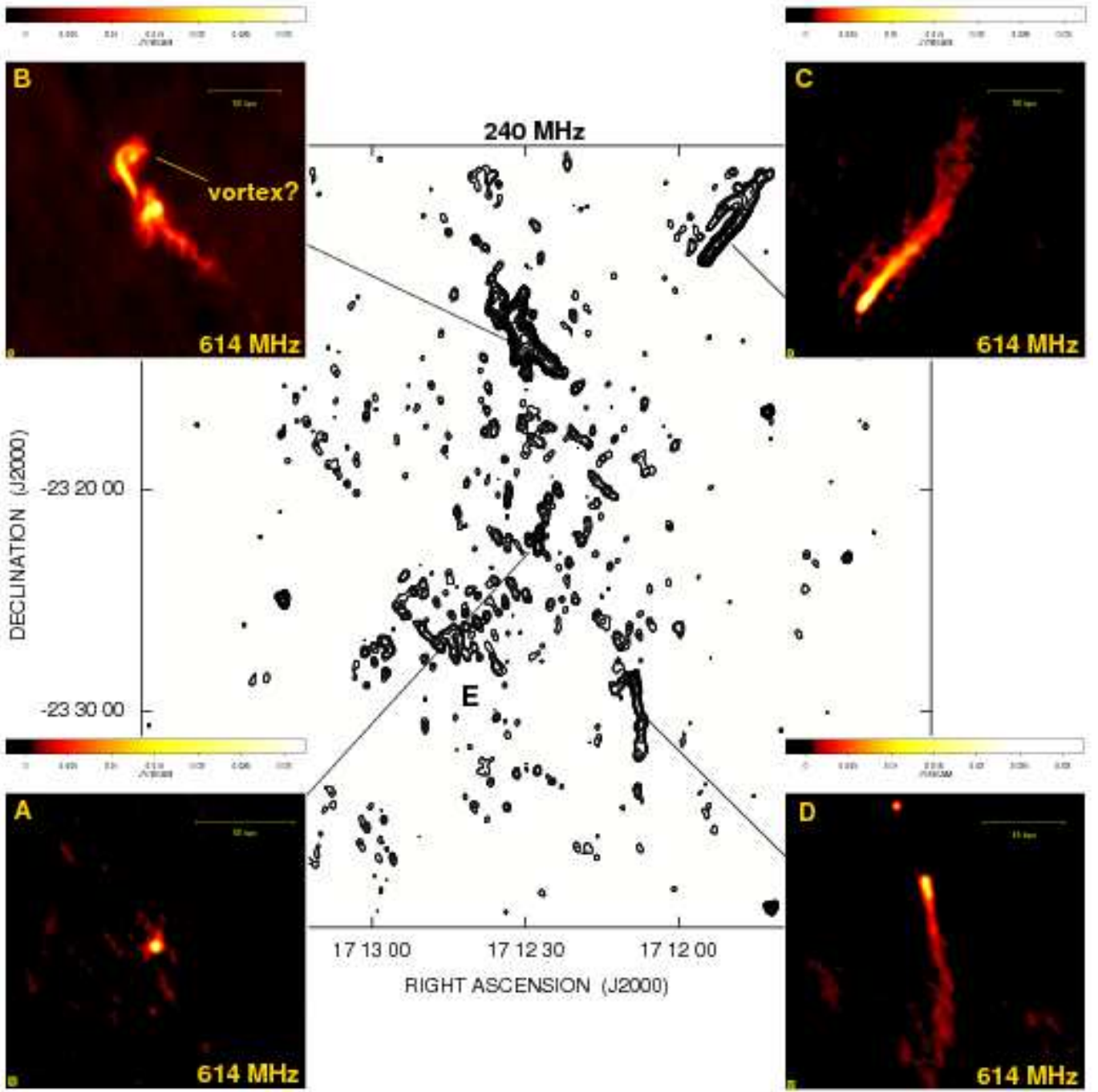}
\caption{Central panel: GMRT radio iso-contours at 240 MHz of the Ophiuchus galaxy cluster at full-resolution. The radio image has a FWHM beam of $18.7\arcsec \times15.9\arcsec$ with PA=$-25.7$\degr and a noise level of 1.1 mJy/beam (1$\sigma$). Contours start at 3.3 mJy/beam ($3\sigma$) and scale by a factor of $\sqrt{2}$. The insets show the GMRT image at 614 MHz of sources A, B, C, and D. The 614 MHz has  a FWHM beam of $7\arcsec \times7\arcsec$ and a noise level of 0.25 mJy/beam (1$\sigma$). }
\label{fig2}
\end{figure*}

We first calibrated the 153 MHz observations from the two runs independently 
 and then we combined the two datasets with task DBCON, 
 performing a final self-calibration run.  In the left panel of Fig.\,\ref{fig1}, we 
present the radio image at 153 MHz resulting from the combination of the 
data sets from the two observing days. The radio image has a FWHM beam 
of $37.1\arcsec \times22.6\arcsec$ with PA=35.1\degr and a noise level 
of 5 mJy/beam (1$\sigma$). 

\subsection{240 MHz} 

The 240 MHz observations consist of two data sets of 4.6 and 4.3 hours on 
source, taken on 23 and 28 August 2008, respectively. The August 23 
observations were performed for both the LSB and the USB with a total 
bandwidth of 8 MHz splitted in 128 channels of 62.5 kHz in width. The 
August 28 observations were performed for the USB with a total bandwidth 
of 8 MHz splitted in 64 channels of 125 kHz in width.  The flux density 
scale was calibrated by the source 3C286. The source 1830-360 
was observed at intervals of $\sim30$ minutes and used as secondary 
phase and gain calibrator. The 
bandpass was calibrated using the sources 3C286 and 1830-360.  Task 
FLGIT was applied to the whole dataset and about 26\% of the data were 
eliminated because of RFIs. Both data sets were averaged in frequency to 
6 channels of 1 MHz in width in order to reduce noise still keeping the 
bandwidth smearing effect under control. We processed the two data set
 separately applying several cycles of 
imaging and self-calibration. The datasets were then 
combined together with task DBCON and we performed a final self-calibration run. 
The contour levels of the full-resolution 240 MHz image are shown in the 
central panel of Fig.\,\ref{fig2}. The radio image has a FWHM beam of 
$18.7\arcsec \times15.9\arcsec$ with PA=-$25.7$\degr and a noise level of 
1.1 mJy/beam (1$\sigma$).

\begin{figure*}[t]
\centering
\includegraphics[width=18 cm]{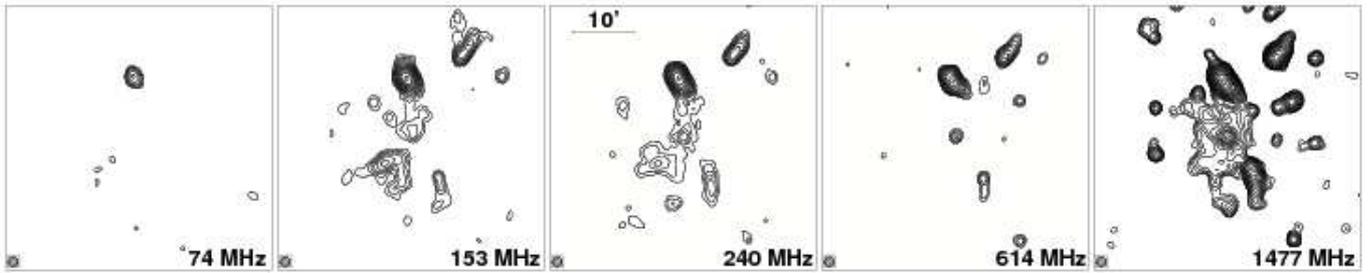}
\caption{Radio images of the Ophiuchus cluster of galaxies at $92\arcsec\times 92\arcsec$ FWHM beam resolution. Contours start at 3$\sigma$ level and 
scale by a factor of $\sqrt{2}$. Left to right: 74 MHz (VLSS; $\sigma=280$ mJy/beam), 153 MHz (GMRT; $\sigma=18$ mJy/beam),
 240 MHz (GMRT; $\sigma=10$ mJy/beam), 614 MHz (GMRT; $\sigma=5.5$ mJy/beam), 1477 MHz (VLA; $\sigma=0.16$ mJy/beam).}
\label{fig3}
\end{figure*}

\subsection{614 MHz} 

The observations were performed on 28 August 2008 for a total of 4.3 
hours on source. Data were recorded for both the USB and the LSB with a 
16 MHz bandwidth splitted in 128 channels of 125 kHz in width. The 
central frequencies of the USB and LSB are 606 and 622 MHz, 
respectively.  The flux density scale was calibrated by using the 
source 3C286. The phase 1830-360 
was observed at intervals of $\sim30$ minutes. The bandpass was 
calibrated using the sources 3C286 and 1830-360. In order to improve the 
signal to noise ratio, data were averaged in frequency by collapsing the 
bandwidth to 6 spectral channels of 2 MHz in width. The 6-channels 
dataset was carefully inspected in order to excise the RFIs. Several 
cycles of imaging and self-calibration were applied to remove the 
residual phase variations. 

The USB and the LSB were imaged separately and then averaged 
together to produce a final image at a frequency of 614 MHz with a FWHM 
beam of $7\arcsec \times7\arcsec$ and a noise level of 0.25 mJy/beam 
(1$\sigma$). Cut-outs of the 614 MHz image are shown in the insets of 
Fig.\,\ref{fig2}.

\section{Results}
We analyze the results of the GMRT observations with a particular emphasis
 on the determination of the radio spectrum of the cluster discrete sources and of the mini-halo. 

\subsection{Optical and radio properties}
The Ophiuchus cluster is one of the most luminous X-ray galaxy clusters in the local Universe, but
 its optical properties are not very well known because of its
unfortunate line-of-sight. The cluster lies in projection at only about 10\degr~ from the Galactic Centre and hence
it is highly obscured. 
Recently, Ophiuchus has been studied in the optical band
by Wakamatsu et al. (2005) who derived spectroscopic redshifts for about 200 galaxies
to within 5\degr~from the cD galaxy at the centre of the cluster core.
The velocity dispersion of the Ophiuchus cluster is found to be $1050\pm50$ km/s. Such a 
large velocity dispersion is consistent with its large X-ray luminosity. Moreover, several 
clusters and groups of galaxies are observed to within a distance of 8\degr~from the cluster centre, 
indicating that the Ophiuchus concentration may be a supercluster comparable in richness
to the Coma$-$A1367 system, as early suggested by Djorgovski et al. (1990).

In the top left panel Fig.\,\ref{fig1}, the overlay of the GMRT radio 
iso-contours at 153 MHz to the ROSAT PSPC X-ray image of the Ophiuchus 
galaxy cluster is shown. The X-ray image is in the 0.1$-$2.4 keV energy band and 
it has been background subtracted, divided by the exposure image,
 and smoothed with a Gaussian kernel with $\sigma=30\arcsec$.

\begin{figure*}[t]
\centering
\includegraphics[width=18 cm]{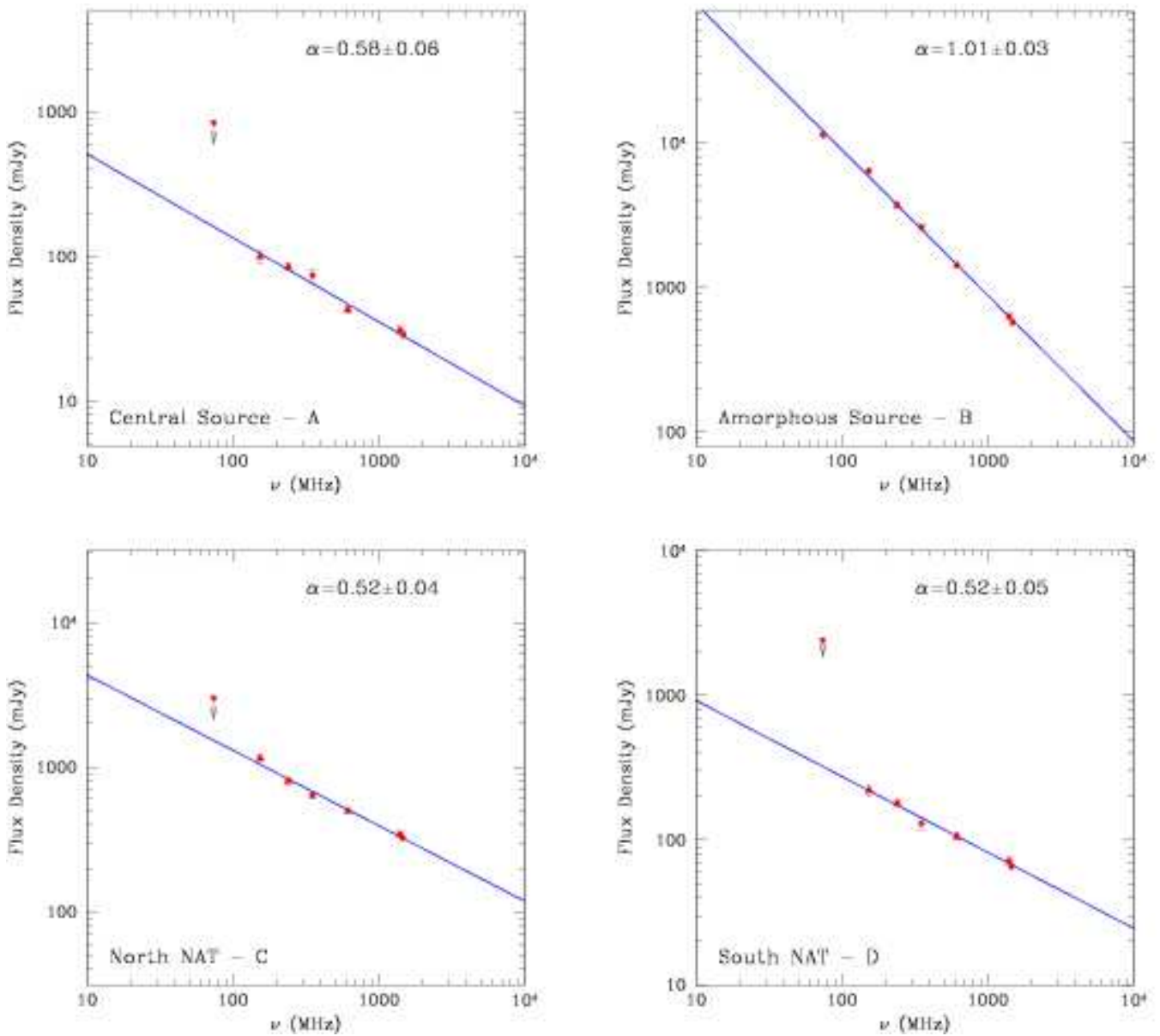}
\caption{Integrated radio spectra for the central source A (top-left), for the amorphous source B (top-right), and for the two tailed sources C and D 
(bottom panels). The triangles represent the GMRT flux density measurement from this work while the dots are the data taken from the literature. The lines represent the best fit of a power law with index $\alpha$. }
\label{fig4}
\end{figure*}

The most relevant radio features we detected are 
labeled A through E. At the centre of the cluster we detected a faint 
point-source (labeled A) whose position is coincident with the peak of 
the X-ray emission. This is the same point source detected with the VLA 
at 1477 MHz by Govoni et al. (2009), see Fig.\,\ref{fig1} top right panel.
Radio source A is associated with the prominent cD galaxy which lies at the centre of 
the cluster (Fig.\,\ref{fig1}, bottom left panel).
Source A appears point-like at our resolution 
and sensitivity. The brightest radio source in the field is source B, which is 
located about 10\arcmin~north to the centre. Source B is an extended 
source with an angular size $174\arcsec\times 50\arcsec$, corresponding 
to a projected linear size of about $100\times 30$ kpc (see top-left 
inset in Fig.\,\ref{fig2}). There is no obvious optical identification 
for this radio source, whose rather peculiar morphology makes its 
classification very uncertain. In fact, it lacks of any of the typical 
feature observed in ordinary radio galaxies, like core, jets, or lobes. 
Source B is neither a head-tail radio galaxy. Rather, the source has an 
amorphous filamentary structure. The south-east part is characterized by 
several threads of radio emission emerging perpendicular to the source's 
major axis.  The north-west part of the source is composed by a single 
filamentary feature whose tip bend backwards to form what resemble a 
vortex-like structure. Instead, sources C and D are the typical cluster 
tailed radio galaxies. Source C is about 300\arcsec~long (165 kpc) and 
is pointing north-east to south-west. Source D is faint tail about 
230\arcsec~long (127 kpc) pointing from south to north. 
The apparent difference in radial velocity between sources C and D and 
the Ophiuchus cluster is $\gtrsim$1000 km/s. We assume here a mean 
recession velocity of 9063 km/s for the central region of the 
Ophiuchus cluster (Wakamatsu et al. 2005), and that the two radio 
sources are associated to galaxies 2MASX J17115542-2309423 (c$z$ = 8050 km/s) 
and 2MASX J17120908-2328263 (c$z$ = 7469 km/s) (Hasegawa et al. 2000). 
These properties are further confirmed by the spectral aging analysis presented in Sect.\ref{aging} 
and indicate that the ram-pressure model can easily explain the radio jet deflection. 
The two galaxies could have such a 
high velocity with respect to the ICM either because they are infalling 
individually towards the cluster centre or because they are part of 
merging sub-clusters. The merger induced bulk motion of the galaxies in 
the ICM would then be responsible for bending the radio jets. The latter 
scenario would be in agreement with the results by Bliton et al. (1998), 
who derived that narrow-angle tailed radio galaxies are preferentially 
found in dynamically complex clusters.

\begin{figure*}[t]
\centering
\resizebox{\hsize}{!}{\includegraphics{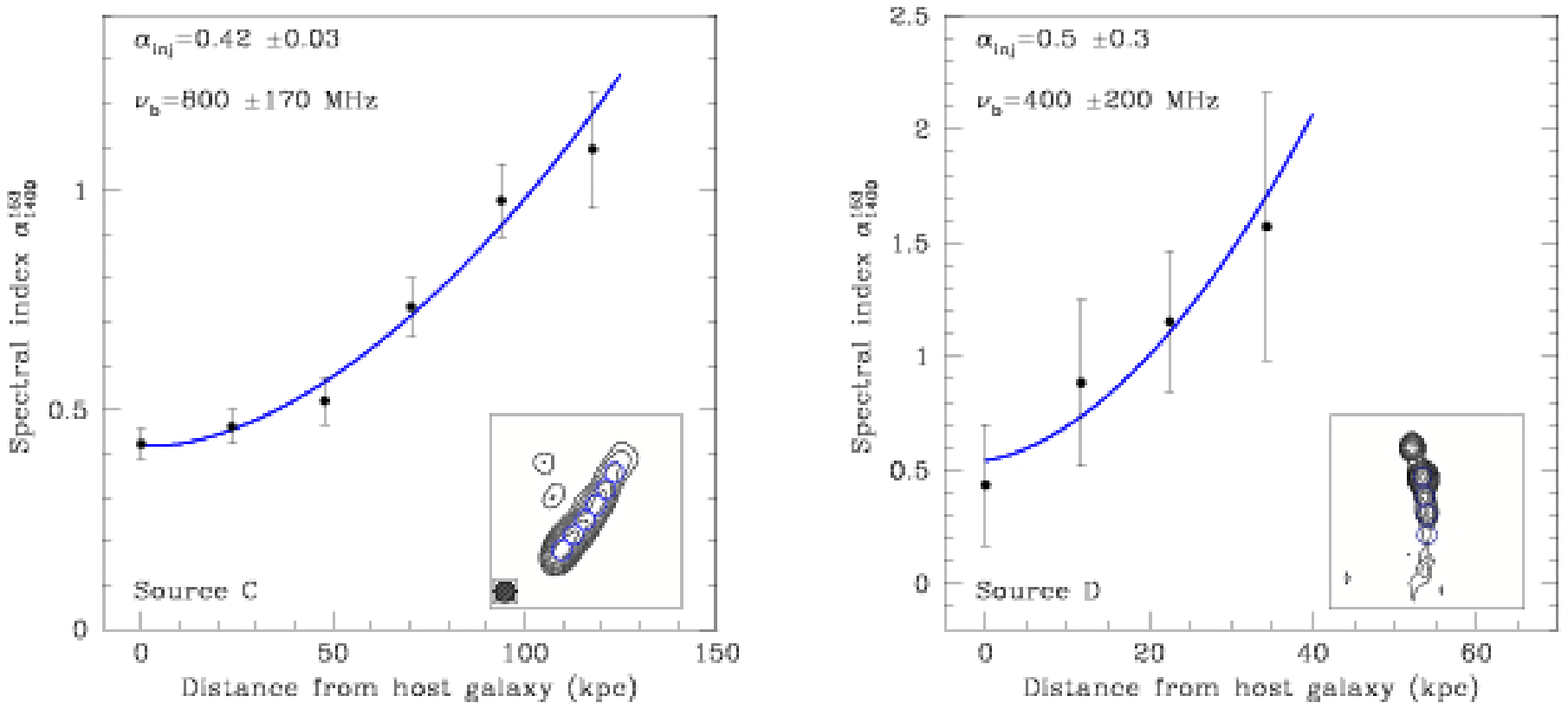}}
\caption{Spectral index profile between 153 and 1400 MHz along the tail for radio sources C (left panel) and D (right panel). Data points have been sampled in circular boxes of one beam-width,
as shown in the inset. The line is the expected spectral index trend for the case of constant advance speed.}
\label{fig5}
\end{figure*}

Finally, we detected an extended patch of diffuse emission at about 
5\arcmin~south-west from the cluster centre. This feature, labelled E in 
Fig.\,\ref{fig1}, is the only part of the mini-halo visible in the 153 
MHz GMRT image at full-resolution. This patch is also present in the 240 
MHz image at full-resolution shown in Fig.\,\ref{fig2}, where few other 
``fragments'' of the mini-halo can be observed all around the cluster 
centre. At 614 MHz the diffuse emission of the mini-halo is too faint to 
be detected at full-resolution. 

The VLA image at 1477 MHz shown in the top-right panel of Fig.\,\ref{fig1} has a 
very good sensitivity to the extended emission. Given the noise level
of about 0.1 mJy and the relatively large resolution of $91\arcsec \times 40\arcsec$
provided by the VLA in D configuration, the $3\sigma$ sensitivity level on the
mini-halo diffuse emission is of $0.07 ~\mu$Jy/arcsec$^2$. For comparison,
 the sensitivities of the full-resolution GMRT images are of 15.8, 8.9, and 13.5 $\mu$Jy/arcsec$^2$
 at 153, 240, and 614 MHz. It should be considered that the observed surface brightness at 1477 MHz 
of the mini-halo is at best of $\lesssim 0.5 ~\mu$Jy/arcsec$^2$ and hence the
minimum spectral index required to detect the mini-halo at 153 and 240 MHz is $\alpha> 1.5$.
The detection turns out to be prohibitive at 614 MHz since the required mini-halo spectral index should be as 
high as $\alpha> 3.8$.

Thus, in order to improve the signal to noise ratio of the GMRT data, 
we realized a set of images with natural weighting (ROBUST=5 in IMAGR) and by tapering the 
longest-baseline in order to smooth the angular resolution to $92\arcsec \times 
92\arcsec$. This is the resolution adopted in the VLA study of the mini-halo
at 1477 MHz by Murgia et al. (2009). The radio iso-contours of these
 images are presented in Fig.\,\ref{fig3}. 
The degradation of the resolution of the images results in an improved sensitivity to the diffuse 
emission. At $92\arcsec \times 92\arcsec$ resolution,
 the $3\sigma$ sensitivities of the GMRT images are of 5.6, 3.1, and 1.6 $\mu$Jy/arcsec$^2$
respectively at 153, 240, and 614 MHz.

The mini-halo is clearly detected at 153 and 240 MHz, the frequencies at
which we reached the best sensitivity to the low-surface brightness
diffuse emission, while it is too faint to be 
detected at 614 MHz even at $92\arcsec \times 92\arcsec$ resolution.
The most prominent feature of the mini-halo at low frequencies is still the patch E, which is marginally 
visible also in the VLSS at 74 MHz. However, the higher signal-to-noise ratio achieved in the 153 and 240 MHz 
images at low-resolution permits to reveal a bridge of diffuse radio 
emission which is aligned along the cluster major axis and connects the 
patch E, the central source A, and the amorphous source B.

\subsection{Integrated radio spectra of discrete sources.}

Determining the radio spectra of the discrete sources is important in order 
to understand their nature and, possibly, their connection with the 
mini-halo.  We analyzed the integrated radio spectra of sources A, B, C, and D 
by complementing the GMRT measurements at 153, 240 and 614 MHz with the 
flux densities available in literature. In particular, we made use of 
the VLSS at 74 MHz, the WISH survey at 325 MHz, the NVSS at 1400 MHz and 
the VLA 1477 MHz image by Govoni et al. (2009). The flux density of 
source A at the different frequencies has been determined by using AIPS 
task JMFIT. Since sources B, C, and D are extended, we determined the flux density 
by integrating their radio brightness down to the 
3$\sigma$ isophote. All the VLSS flux densities at 74 MHz reported in this
work have been  corrected for the clean bias, following the prescription of 
Cohen et al. (2007). The integrated spectra are 
reported in Table\,\ref{tab2}. The flux density uncertainties include a 
5\% absolute calibration error.

In Fig.\,\ref{fig4}, we present the plots of the integrated radio 
spectra along with a power law fit to the data.  Overall, the GMRT 
measurements are in agreement both with the fit and with the adjacent
 data points taken from the literature, thus 
providing a positive check of the flux density scale in our images. 

Source A is a compact source with a steep spectrum. The spectral index 
of the central source in the considered frequency window is  
$\alpha\simeq 0.6$, a typical value for radio sources.

The amorphous source B presents a power-law radio spectrum with a spectral 
index of about $\alpha=1.01 \pm 0.03$. This is a 
quite usual value for active radio galaxies which makes the interpretation of this object even 
more puzzling. In fact, although the distorted morphology of this radio 
source recalls that of extreme relic sources in clusters of galaxies (see 
e.g. Slee et al. 2001), on the basis of its radio spectrum it cannot be 
classified as an ultra-steep spectrum source. A 5 GHz archive VLA image 
 (not shown) suggests the presence of very weak point-like 
source but the absence of kpc-scale jets. Indeed, it is not clear which 
mechanism is powering the relativistic electrons in this extended radio 
source. A possibility could be that source B is a relic radio source
revived by the adiabatic compression caused by a shock wave or a bulk 
gas motion propagating thought the ICM (En{\ss}lin \& Gopal-Krishna 2001). 
However, no particular X-ray feature is visible in coincidence 
of the radio source neither in the ROSAT nor in the \xmm~ images and 
hence the origin of this peculiar radio source as well as its possible
relation with the mini-halo remain, at the moment, unclear.

{Finally, the tailed sources C and D have similar integrated spectra well described by
 a power law with index $\alpha\simeq 0.5$.

\subsection{Spectral aging analysis of sources C and D}
\label{aging}

Sources C and D can be classified as narrow-angle-tails (NATs). This morphology is indicative of a 
strong relative velocity between the host galaxy and the ICM. The ram pressure exerted by the external
 gas bends the radio jets that merge together forming the characteristic tail of radio plasma.
The relativistic electrons at the end of the tail must have been deposited first and hence 
their radio spectrum should be steeper since they suffered larger energy losses. 

In Fig.\,\ref{fig5} we present the spectral index profile between 153 MHz (GMRT) and 1400 MHz (NVSS) as a function of distance 
from the host galaxy for sources C and D. Both the GMRT and the NVSS images 
have been convolved to the same resolution of $45\arcsec\times 45\arcsec$ and re-gridded to a common pixel size 
in order to be properly compared. 

For both sources, the spectral index increases systematically with 
the increasing distance from the host galaxy, a typical behaviour observed in many NATs.
The spectral steepening can be interpreted in terms of radiative losses of the relativistic electrons.
In particular, we assumed that the radio spectrum is described by a JP model (Jaffe \& Perola 1972)
characterized by a low-frequency zero-age power law with index $\alpha_{inj}$ and an exponential cut off
beyond a high-frequency break, $\nu_{b}$.
The break frequency is related to the radiative age of the relativistic electron, $t_{rad}$, and to the 
source's magnetic field, $B$, through

\begin{equation}
t_{\rm rad}= 1590 \frac{B^{0.5}}{(B^2+B_{\rm CMB}^2) \sqrt{(1+z) \nu_{b}}}   ~~~\rm (Myrs)
\label{age}
\end{equation}

where the magnetic field is in $\mu G$ and the break frequency in GHz while 
$B_{\rm CMB}=3.25(1+z)^{2}\,\mu$G is a virtual magnetic field whose energy
 density equal that of the CMB and accounts for the inverse Compton losses (see also Sect. 4.2).

Following Parma et al. (1999), we assume that the radio plasma in the tail separates from the host galaxy at a constant speed.
In this case, the break frequency scales $\nu_{b}\propto 1/d^2$, where $d$ is the distance along the tail.
Given this trend for $\nu_{b}$, we can compute the expected spectral index on the basis of the JP model as a function of the distance $d$.
Close to the host galaxy ($d\rightarrow 0$) the break frequency $\nu_{b}\rightarrow\infty$; the radio spectrum is a power law with index
 $\alpha_{inj}$. At increasing distance from the host galaxy, $\nu_{b}$ shifts to low frequency an the radio spectrum steepens. 
By fitting  the observed spectral index profile, we derived the injection spectral index and the lowest value for the break frequency. 
For source C  we find $\alpha_{inj}\simeq 0.42\pm 0.03$ and a minimum break frequency of $\nu_{b}\simeq 800 \pm 170$ MHz at
 a distance of 120 kpc from the host galaxy. By using standard formulas (e.g. Pacholczyk 1970),
we estimate the minimum energy magnetic field, $B_{min}$, of the tail by intregrating the radio luminosity from 100 MHz to 10 GHz
 and by assuming a ratio between the energy density of relativistic protons to that of the electrons of $k=1$.
For source C we obtain $B_{min}\simeq 6.7$ $\mu$G and, on the basis of Eq.\,\ref{age}, we calculate a radiative age of $t_{\rm rad}=80 \pm 10$ Myr. 
The corresponding advancing speed of the tail is $v\gtrsim 1460\pm 160$ km/s, in agreement with the dispersion velocity of 
galaxies in the Ophiuchus cluster.

Source D is fainter than source C and thus the uncertainties on the best fit parameters are larger.  However, 
we find $\alpha_{inj}\simeq 0.5\pm 0.3$ and a lowest break frequency of $\nu_{b}\simeq 400 \pm 200$ MHz at
a distance of about 40 kpc from the host galaxy. We calculate for the radio source a minimum energy magnetic field of 
$B_{min}\simeq 8.4$ $\mu$G, and we estimate a radiative age of $t_{\rm rad}=87 \pm 26$ Myr which corresponds to an advancing speed for 
the tail of $v\gtrsim 450\pm120$ km/s, i.e. smaller than that of source C.

It is worth noting that the estimated advancing velocities should be regarded as lower limits if the tail's length has been  
 significantly shortened by projection effects.

\subsection{The mini-halo spectral index image}

The main goal of this work is to constrain the spectral index of the 
mini-halo in the Ophiuchus cluster. This is a particularly hard task 
since the mini-halo is very faint and extended. The best compromise 
between sensitivity and resolution is obtained in the 240 MHz image
at $92\arcsec\times 92\arcsec$ resolution.  
In the top-left panel of Fig.\,\ref{fig6}, we present the spectral 
index image between 240 and 1477 MHz with the 240 MHz radio iso-contours overlaid.
The spectral index image is calculated only on those pixels whose brightness
is above the $3\sigma$ level at both frequencies.
The overall radio spectrum of the mini-halo is steep. The spectral index 
ranges from about  $\alpha\simeq 1.5$, close to the cluster centre, up to about
 $\alpha\simeq 2.0$ in correspondence of patch E. 
In the top-right panel of Fig.\,\ref{fig6} we show the spectral index uncertainty 
which is in the range from 0.05 to 0.2.

Bottom panels of Fig.\,\ref{fig6} show the radio spectrum of the 
mini-halo at four different sample positions.
In addition to the GMRT data, the spectra also include the measurements obtained from the 
VLSS at 74 MHz and the VLA at 1477 MHz. All the considered images 
have been re-gridded to a common geometry and convolved at the same angular 
resolution of $92\arcsec\times 92\arcsec$.
The radio spectra between 74 and 1477 MHz indicate no deviation from a
power-law model, although this could be due to the 
comparatively small frequency range considered.

\begin{table}
\caption[]{Integrated spectra for cluster discrete sources.}
\begin{center}
\begin{tabular}{cccc}
\hline
%\noalign{\smallskip}
Source   & $\nu$  & Flux Density & Reference\\
         &      MHz  &   mJy     &    \\
\noalign{\smallskip}
\hline
\noalign{\smallskip}
Central Source -A     &  74    & $< 840$  & e \\ 
                &  153 & $100 \pm 11$  & a \\ 
                &  240   & $ 85  \pm 4$   & a \\
                &  352   & $ 74  \pm 8 $   & b  \\
                &  614   & $ 43  \pm 2  $   & a  \\
                &  1400  & $ 31  \pm 2  $   & c  \\
                &  1477  & $ 29  \pm 2  $  & d  \\
\hline
\noalign{\smallskip}
Amorphous source -  B            &  74    & $11330 \pm 570$  & e \\ 
                &  153   & $6380  \pm 320$  & a\\ 
                &  240   & $3700  \pm 190$  & a\\
                &  352   & $2610  \pm 130$  & b\\
                &  614   & $1420  \pm 70 $  & a\\
                &  1400  & $632   \pm 32 $  & c\\
                &  1477  & $574   \pm 29 $  & d \\

\hline
\noalign{\smallskip}
North NAT source -  C    &  74    & $<3000$  & e \\      
                & 153    & $1150  \pm 60$  & a\\ 
                &  240   & $800  \pm 40$  & a\\
                &  352   & $636  \pm 32$  & b\\
                &  614   & $497  \pm 24 $  & a\\
                &  1400  & $343   \pm 17 $  & c\\
                &  1477  & $326   \pm 16 $  & d \\
\hline
\noalign{\smallskip}
South NAT source -  D    &  74    & $<2400$  & e \\           
                & 153    & $220  \pm 20$  & a\\ 
                &  240   & $180  \pm 11$  & a\\
                &  352   & $130  \pm 12$  & b\\
                &  614   & $106  \pm 6 $  & a\\
                &  1400  & $72   \pm 4 $  & c\\
                &  1477  & $66   \pm 3 $  & d \\
\hline
\multicolumn{4}{l}{\scriptsize a) GMRT, this work; b) WISH survey, De Breuck et al. (2002);}\\
\multicolumn{4}{l}{\scriptsize c) NVSS, Condon et al. (1998); d) VLA, Govoni et al. (2009);}\\
\multicolumn{4}{l}{\scriptsize e) VLSS, Cohen et al. (2007).}
\label{tab2}
\end{tabular}
\end{center}
\label{integratedspectra}
\end{table}

Patch E is detected also at 74 MHz in the VLSS and has a spectral index of $\alpha\simeq 2.04\pm0.09$ 
(see Fig.\,\ref{fig6}, bottom-left spectrum).
Patch E appears as an isolated feature at a comparatively large distance from the center of the cluster. However, 
it is important to stress again that the VLA image at 1477 MHz has a much higher 
dynamic range with respect to the GMRT image at 240 MHz. Indeed, there are
regions of the mini-halo which are clearly detected at 1477 MHz but that are not represented in Fig.\,\ref{fig6} 
simply because their spectrum is not steep enough.

\subsection{Azimuthally averaged radial profile of the mini-halo emission}

The GMRT images at 153 and 240 MHz at $92\arcsec \times 92\arcsec$ resolution
have enough sensitivity to allow us the analysis of the azimuthally averaged radial 
profiles of the mini-halo emission.
The surface brightness of mini-halos in clusters of galaxies decreases with 
increasing distance from the cluster center, eventually falling below 
the noise level of the radio images.  Although deviations of the diffuse 
emission from spherical symmetry are often observed, the azimuthally 
averaged radial profiles are indeed quite smooth and regular. We derived the 
azimuthally averaged brightness of the Ophiuchus mini-halo at 153 and 240 MHz 
and we compare the result with the finding obtained with the VLA at 1477 
MHz by Murgia et al. (2009).

\begin{figure*}[t]
\centering
\includegraphics[width=14 cm]{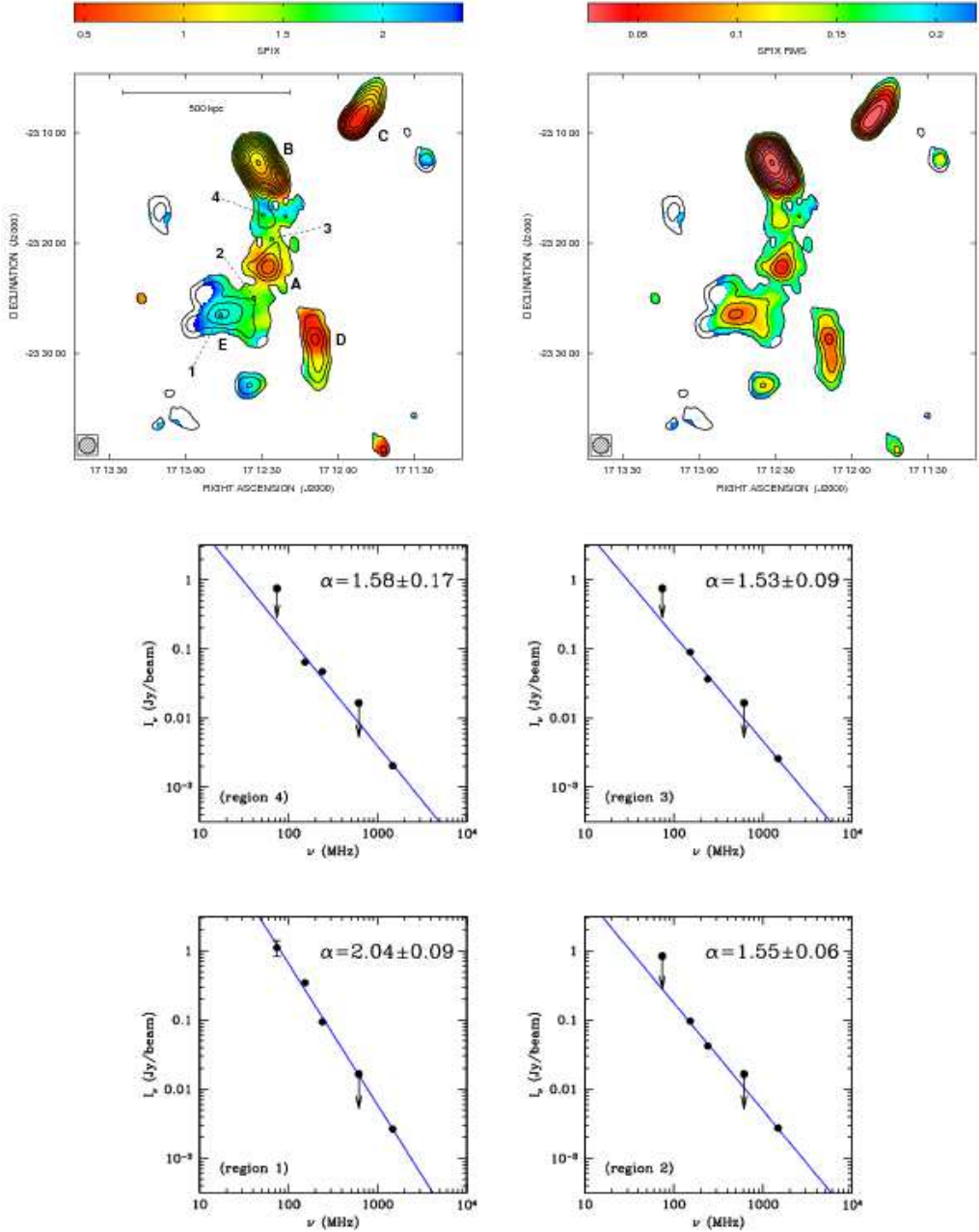}
\caption{Top left panel shows the spectral index image of the mini-halo in the Ophiuchus cluster between 240 and 1477 MHz at an angular resolution of $92\arcsec \times 92\arcsec$. The 
iso-contours represent the 240 MHz image obtained by tapering the long baselines between the GMRT antennas. Contours starts at 30 mJy/beam (3$\sigma$) and increase
by a factor of $\sqrt{2}$. The 1477 MHz image is taken from Govoni et al. (2009) and has a noise level of 0.16 mJy/beam (1$\sigma$). 
The spectral index is calculated only in those pixels whose brightness is above 3$\sigma$ at both frequencies. Top right panel shows the image of the spectral index uncertainty. Bottom panels show the mini-halo spectrum between 74 MHz and 1477 MHz in the four sample positions indicated in the top right panel.}
\label{fig6}
\end{figure*}

In the top panel of Fig.\ref{fig7} we show the azimuthally averaged 
radio halo brightness profiles obtained from the 153, 240 and 1477 MHz images 
at 92\arcsec~resolution. 
Each data point represents the average 
brightness in concentric annuli of half beam width centered on the X-ray 
peak. Discrete sources have been masked out and 
excluded from the statistics. We considered only data points whose observed brightness
is five times above the error on the radial average.
Following Murgia et al. (2009), in order 
to carefully separate the contribution of the mini-halo from that of the 
central radio galaxy, we fitted the total brightness profiles with a 
central point source plus the radio mini-halo diffuse emission

\begin{equation}
 I(r)=I_{PS}(r)+I_{MH}(r).
\label{expodisk+psf}
\end{equation}

In particular the profile of the central point source has been
characterized by a Gaussian of the form
\begin{equation}\label{expfit}
 I_{PS}(r)=I_{0_{PS}}\,e^{-(r^2/2\sigma_{PS}^2)}
\label{psf}
\end{equation}
while the brightness profile of the mini-halo has been characterized by an
exponential law of the form
\begin{equation}\label{expfit}
I_{MH}(r)=I_{0_{MH}}\,e^{-r/r_{e}}.
\label{expodisk}
\end{equation}

The best fit of the total model in Eq.\,\ref{expodisk+psf} is 
represented by the solid lines in the top panel of Fig.\ref{fig7}.  The contribution from 
the mini-halo exponential disk alone is represented by the dotted lines.  
The fit is performed in the image plane as described in Murgia et al. 
(2009). In order to properly take into account the resolution, the 
exponential model is first calculated in a 2-dimensional image, with the 
same pixel size and field of view as that observed, and then convolved 
with the same beam by means of a Fast Fourier Transform. The resulting 
image is masked exactly in the same regions as the observations. 
Finally, the model is azimuthally averaged with the same set of annuli 
used to obtain the observed radial profile. All these functions are 
performed at each step during the fit procedure.  As a result, the 
values of the central brightness, $I_{0_{MH}}$, and the e-folding radius 
$r_{e}$ provided by the fit are deconvolved quantities and their 
estimate includes all the uncertainties related to the masked regions 
and to the sampling of the radial profile in annuli of finite width.  
The best fit parameters are reported in Table\,\ref{tab3}.

At 153 MHz the best fit of the exponential model yields a central 
brightness of $I_{0_{MH}}\simeq 7.0\,\mu$Jy/arcsec$^2$ and $r_{e}\simeq 510$ kpc while 
at 240 MHz the best fit yields $I_{0_{MH}}\simeq 3.9\,\mu$Jy/arcsec$^2$ and $r_{e}\simeq 450$ 
kpc.

In the bottom panels of Fig.\ref{fig7} we trace the spectral index radial 
profile of the mini-halo between 153 and 240 MHz and between 240 and 1477 MHz.
The radial profile of the low-frequency spectral index between 153 and 240 MHz is fairly 
constant to a value of about $\alpha\simeq 1.3 \div 1.5$.
On the other hand, the e-folding radius of the mini halo at low frequency is more than four times larger 
than at 1477 MHz. This results in a progressive steepening of the 
spectral index with the increase of the distance from the cluster 
centre. The spectral index between 240 and 1477 MHz increases steadily from $\alpha\simeq 1.3$, at the 
cluster centre, up to  $\alpha\simeq 2.0$, at the mini-halo periphery.

We calculated the model flux densities in an area with a radius of $r=7\arcmin$ (corresponding to about 230 kpc) 
from the cluster centre. This is the extraction region of the HXR emission
used by Nevalainen et al. (2009).
The mini-halo flux densities have been obtained by the formula (Murgia et al. 2009):

\begin{equation}
 S_{MH}= 2\pi \int_{0}^{r^{\prime}} I_{MH}(r) r dr=2\pi [1+e^{-r^{\prime}/r_e}(-r^{\prime}/r_e-1)] \cdot r_e^2 I_{0_{MH}}
\end{equation}
where we set $r^{\prime}=7\arcmin$.

We obtained: $S_{153}=2900$ mJy, $S_{240}=1560 $ mJy, and $S_{1477}=85$ mJy.
The mini-halo is not detected at 74 and 614 MHz, the upper limit to the flux density
 in a 7\arcmin~area are of $S_{74}<48500$ mJy and  $S_{614}<900$ mJy, respectively. These limits have been
calculated by imposing that the average surface brightness of the diffuse emission
 is lower than the $3\sigma$ noise level of the corresponding radio image.
  
The global radio spectrum of the mini-halo is shown in Fig.\ref{sed}. The low and high frequency spectral indices are 
respectively of $\alpha_{240}^{153}\simeq 1.38\pm 0.27$ and $\alpha_{1477}^{240}\simeq 1.60 \pm 0.05$. 
Indeed, a hint of spectral steepening is seen at high frequency, although the two values are still compatible to within the errors.

It is worthwhile to note that the global low-frequency radio spectral index 
 is fully consistent with the reported range of 1.2$-$1.5 obtained in the \xmm/\intgr~ analysis
 by Nevalainen et al. (2009), see next Section for further details.

\begin{table}
\caption{Properties of the Ophiuchus mini halo derived from the fit procedure.}
\begin{center}
\begin{tabular} {ccccccc}
\hline
$\nu$ &     $r_e$    &$r_e$   &  $I_{0_{MH}}$      & $S_{MH}(r\le7\arcmin)$  &$\chi^2/d.o.f.$ \\
  MHz     &    arcsec &kpc     & $\mu$Jy/\arcsec$^2$  & mJy  &        \\
\hline\\
\vspace{0.2cm}

153  & $930^{+1500}_{-340}$  &$510^{+825}_{-190}$ & $7.0^{+1.7}_{-1.5}$& $2900^{+250}_{-250}$ &0.66\\
\vspace{0.2cm}

240  & $810^{+1620}_{-330}$  &$450^{+890}_{-180}$ & $3.9^{+1.1}_{-0.9}$& $1560^{+140}_{-140}$ & 0.46\\
\vspace{0.2cm}

1477 & $191^{+23}_{-19}$   &$105^{+13}_{-11}$&$0.58^{+0.08}_{-0.07}$& $85^{+3}_{-3}$ &2.9\\
\hline

\end{tabular}
\label{tab3}
\end{center}
\end{table}

\section{Discussion}

The Ophiuchus galaxy cluster is one of the very rare clusters where the 
non-thermal component is revealed both in the radio and X-ray bands.

Assuming that the X-ray component arises from inverse-Compton (IC) 
scattering from the non-thermal electrons with the photons of the CMB,
the volume-averaged cluster magnetic field, $B_V$, can be derived 
essentially from the ratio between the power emitted through synchrotron and IC
(see e.g. Blumenthal \& Gould 1970).

Using \xmm\ and \intgr\ spectra of the cluster, and fixing the spectral 
index of the power law to 1.4 (this work), we extracted the fluxes of 
the non-thermal component in 5 different X-ray energy bands (0.6$-$2, 2$-$5, 5$-$10, 
20$-$40, and 40$-$80 keV). For the details of the data analysis and the 
modelling of the different thermal components, we refer to
Nevalainen et al. (2009). The total non-thermal flux in the 20$-$80 keV band is 
$5.5\times10^{-12}$ ergs $s^{-1}$ cm$^{-2}$, in agreement with the upper 
limits derived from \emph{Swift} (Ajello et al. 2009) and \emph{Suzaku} 
(Fujita et al. 2008) data. 

Using these measurements together with the radio fluxes presented in 
Table 3, we constructed a Spectral Energy Distribution (SED) of the 
non-thermal emission (see Fig. \ref{sed}). We also added to the SED the 
upper limits at 74 MHz (VLSS) and 607 MHz (GMRT), as 
well as the \emph{Swift} upper limit (Ajello et al. 2009).

\subsection{SED modeling}

To model the SED, we used the exact derivation of the synchrotron and IC 
spectrum for a single electron from Blumenthal \& Gould (1970), and numerically 
convolved the resulting spectrum with the differential distribution of 
non-thermal electrons $N(\gamma,\theta)$. Here $\gamma$ is the electron's Lorentz factor while $\theta$ is the 
pitch-angle between its velocity and the local direction of the magnetic 
field. Indeed,  $dN=N(\gamma,\theta)d\gamma d\theta$ is the number 
density of non-thermal electrons with Lorentz factor between
 $\gamma$ and $\gamma+d\gamma$ and pitch angle between  $\theta$ and $\theta+d\theta$.

We assume that the energy spectrum of the non-thermal electrons is described by a power law 
with index $p$, with a high-energy cut-off $m_ec^2 \gamma_{max}$ and a low-energy 
cut-off $m_ec^2\gamma_{min}$:

\begin{equation} 
N(\gamma,\theta)=K_0 \gamma^{-p}(\sin\theta)/2,
\label{Nel}
\end{equation}
where we consider an isotropic distribution of pitch angles.

The radio emissivity as a function of the pitch angle in a uniform
magnetic field of strength $B$ is given by:
\begin{equation} 
j_{syn}(\nu,\theta)=\int_{\gamma_{min}}^{\gamma_{max}}C_f B\sin\theta\,N(\gamma,\theta)F_{syn}(\nu/\nu_{c})d\gamma ~~~{\rm \left(\frac{erg}{cm^{3}s\,Hz}\right)}
\label{jradiotheta}
\end{equation}
where $F_{syn}(\nu/\nu_{c})$ is the synchrotron kernel (see e.g.  Blumenthal \& Gould 1970; Rybicki \& Lightman 1979), while
  
\begin{equation} 
\nu_{c}=4.2\times 10^{-6} B_{\mu G} \sin\theta\,\gamma^2    ~~~\rm (MHz).
\label{nusyn}
\end{equation}
The constant $C_f=2.3444\times 10^{-22}$ erg gauss$^{-1}$ depends only on fundamental
physical constants.

\begin{figure}[t]
\centering
\resizebox{\hsize}{!}{\includegraphics{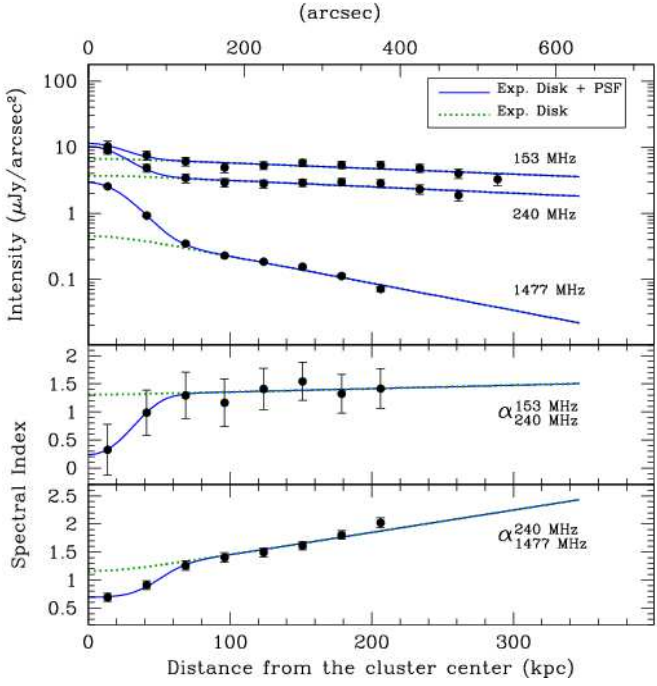}}
\caption{Azimuthally averaged radial profiles of the mini-halo radio intensity at 153, 240 and 1477 MHz (top panel). 
We considered only data points whose observed brightness is five times above the error on the radial average.
The solid lines represent the best fit of the exponential disk model plus the contribution by the central point source.
The dotted lines represent the exponential disk model alone. Mid and bottom panels show the radial profiles of the spectral
 index between 153 and 240 MHz and between 240 and 1477 MHz, respectively.}
\label{fig7}
\end{figure}

Due to the high-beaming of the synchrotron radiation pulses, if the magnetic field direction
makes an angle $\theta^{\prime}$ with respect to the line-of-sight, only relativistic
electrons with pitch angle $\theta\simeq \theta^{\prime}$ will be observed. In this work,
 we suppose that the intra-cluster magnetic field is completely tangled in an infinitesimally
 small scale compared to the mini-halo size. With this assumption, the synchrotron
emissivity averaged over all the possible magnetic field directions\footnote{From simple geometric considerations it follows that 
 the probability to observe an angle $\theta^{\prime}$ between the local direction of a random magnetic field and the line-of-sight is proportional to $\sin \theta^{\prime}$.} is

\begin{equation} 
j_{syn}(\nu)=\int_{0}^{\pi} j_{syn}(\nu,\theta^{\prime}) \frac{\sin\theta^{\prime}}{2} d\theta^{\prime}.
\label{jradio}
\end{equation}

By using standard formulas, we calculated the HXR emission deriving from the inverse Compton scattering of the CMB
 photons by the same population of relativistic electrons in Eq.\,\ref{Nel}:

\begin{equation} 
j_{IC}(\epsilon_1)=\int_{0}^{\infty}\int_{\gamma_{min}}^{\gamma_{max}} N(\gamma) F_{IC}(\epsilon,\gamma,\epsilon_1) d\gamma d\epsilon ~~~{\rm \left(\frac{photons}{cm^{3}s\,erg}\right)}
\label{jic}
\end{equation}
where $F_{IC}(\epsilon,\gamma,\epsilon_1)$ is the kernel emission of a single electron with Lorentz factor $\gamma$, scattering
to energy $\epsilon_1$ a segment of the distribution of CMB photons with initial energy $\epsilon$ (see e.g. Eq.\,2.48 in Blumenthal \& Gould 1970).
Here $N(\gamma)=\int_{0}^{\pi}N(\gamma,\theta) d\theta$.

The synchrotron and inverse Compton emissivities are converted to flux densities by  
multiplying Eq.\,\ref{jradio} and Eq.\,\ref{jic} by a volume of $V=1.51\times10^{72}$ cm$^3$ (that is the 
volume of a sphere whose radius corresponds to a projected distance of $r=7\arcmin$~ from the cluster centre) and
then by dividing by the cluster's luminosity distance, $D_{L}$, i.e. $S_{\nu}=j(\nu) \cdot V/(4\pi D_{L}^2)$.

We then fitted our model to the data and extracted the relevant physical parameters. The 
resulting best-fit model is shown as a line in Fig. \ref{sed}, while
the best-fit parameters are reported in Table\,\ref{tab4} along with their 1$\sigma$ uncertainties.
The two upper limits at 74 and 614 MHz, and that in the 20-60 keV band, have been not considered in the fit since 
they have no influence on the $\chi^2$ statistics.

Unfortunately, the available HXR data do not allow us to constrain precisely the value 
of the low-energy cut-off but only to place an upper limit at $\gamma_{min}\lesssim 700$. 
We indeed decided to fix this parameter to the arbitrary value of $\gamma_{min}=300$,  see Sect.\,\ref{energetics}.

The radio and HXR data are consistent with a slope for the energy spectrum of the relativistic electrons
of $p\simeq 3.8\pm 0.4$ and with a high-energy cut off $\gamma_{max}\simeq 3.8\times 10^4$. The power-law index $p$ is related to the radio spectral index via the relation $p=2\alpha+1$. Indeed, the global mini-halo radio spectrum can be described by a 
low-frequency power law spectral index of $\alpha\simeq 1.4$ followed by a high-frequency break at $\nu_{max}\gtrsim 1900$ MHz.

The derived volume-average magnetic field value $B_V$ is found to be

\begin{equation} 
B_V=0.31^{+0.11}_{-0.07} ~~~(\mu G).
\end{equation}

This estimate depends only weakly on the values of the spectral parameters $\gamma_{min}$, $\gamma_{max}$,
 and $p$. It is much sensitive, however, on the assumptions that the distribution of non-thermal electrons is
isotropic and that the magnetic field is completely disordered over the large volume of space we
considered. We may for instance consider the simpler, but less realistic, situation in which the pitch 
angle is $\theta=90\degr$ for all the electrons with a perfectly ordered magnetic field aligned to the plane of
the sky. In this case, the synchrotron emission is maximized and the magnetic field strength
has to be reduced to $B_V\simeq 0.18\,\mu$G. This lower value for the volume-average magnetic field strength is 
consistent, to within the statistical uncertainties, with that in Nevalainen et al. (2009), who indeed assumed $\theta=90\degr$
 and a perfectly ordered magnetic field.

An important point to discuss is that, given a magnetic field strength of $B\simeq 0.3 \,\mu G$, 
only high-energy electrons with $\gamma> 10^4$ can emit at radio frequencies of 153 MHz and above (see Eq.\,\ref{nusyn}).
On other hand, the HXR emission would be tracing relativistic electrons 
of lower energy, with characteristic Lorentz factors in the range $10^3$ $-$ $10^4$. Indeed, the 
radio and HXR emissions are not tracing exactly the same particles. Nevertheless, the low-frequency spectral 
index measured in this work, $\alpha_{240}^{153}\simeq 1.4\pm0.3$, is consistent with the scenario in which
the energy spectrum of the synchrotron electrons most probably belongs to the extrapolation at higher energies
of the power law energy distribution of the electrons radiating in the HXR band through the inverse Compton process.

\subsection{Energetics and particle life-time}
\label{energetics}
The total pressure from the non-thermal electrons is given by

\begin{equation} 
P_{nt}=u_{el}/3 = mc^2/3 \int_{\gamma_{min}}^{\gamma_{max}} K_0 \gamma^{-p} \gamma d\gamma.
\end{equation}

The non-thermal energy pressure depends critically on the low energy cut off of the 
particle distribution. With the choice of $\gamma_{min}=300$, it results $P_{nt}\sim2.5\times 10^{-12}$ ergs cm$^{-3}$. 
Given that the pressure of the thermal plasma is $P_{ther}\sim9.2\times10^{-11}$ ergs cm$^{-3}$ (Nevalainen et al. 2009),
the non-thermal electrons are responsible for 
$\sim$3\% of the total energy budget, so the dynamics of the gas is 
unaffected by the non-thermal electron population. It is worthwhile to mention that
at energies lower than $\gamma \lesssim 700$ Coulomb collisions with the thermal plasma are the most important
energy loss process for the relativistic electrons in Ophiuchus (see e.g. Sarazin 1999). 
In particular, we calculated that for $\gamma_{min}=300$, the heating rate of the intra-cluster medium
produced by the non-thermal electrons through Coulomb collisions still does not exceeds the 
bremsstrahlung cooling rate of the gas. 
The heating rate produced by the non-thermal electrons begins to dominate over the gas cooling rate if the low energy cut off
is lower than $\gamma_{min}\lesssim 180$ (see Colafrancesco \& Marchegiani 2009 for a detailed discussion). We note that in this limiting 
case, the re-heating of the cluster's cooling core by the non-thermal mini-halo could not be neglected. 

Another important consideration is that the energy density of the relativistic
electrons is four orders of magnitude higher than the energy density of the magnetic field,
$B^2/(8\pi)=3.6\times 10^{-15}$ ergs cm$^{-3}$. 
This result would imply that the
mini-halo is not in a minimum energy condition, which requires instead that
the energy densities of particles and field should be nearly equipartited.
This result holds even in the case where the energy distribution is truncated at a 
$\gamma_{min}=700$, that is the upper limit we can place 
on the basis of the current HXR data.

The radiative life-time of the relativistic electrons can be estimated as 

\begin{equation}
t_{\rm rad}\equiv \frac{\gamma}{\dot\gamma} \simeq  \frac{7.73\times 10^{20}}{(B^2+B_{\rm CMB}^2) \gamma}   ~~~\rm (sec)
\label{synage}
\end{equation}
where the magnetic field is in $\mu$G, while $B_{\rm CMB}=3.25(1+z)^{2}\,\mu$G is a virtual magnetic field whose energy
 density equal that of the CMB and accounts for the inverse Compton losses.

Since $B_{\rm CMB}\gg B$, the radiative losses are dominated by the inverse Compton process which 
cool down the high energy electrons with $\gamma=10^4$  (i.e. those radiating at $\sim 153$ MHz) in a time-scale of 
$t_{rad}\simeq 2\times 10^{8}$ yrs. This is a relatively short time compared to
the diffusion time needed by the relativistic electrons to cross the mini-halo.
If we suppose (Melrose 1968) that the relativistic electrons diffuse at the Alfv{\'e}n speed:
\begin{equation} 
v_A\simeq 70 \, B_{\mu G}\,n_{ther}^{-0.5}   ~~~\rm (km/sec)
\end{equation}
where the thermal gas density is expressed in units of 10$^{-3}$cm$^{-3}$, the characteristic diffusion length of
 the synchrotron electrons radiating at 153 MHz results in just $L_{\rm diff}\sim 4.4$ kpc, i.e.
two orders of magnitude smaller than the mini-halo size. This confirms the well-know results that the non-thermal electrons 
 must be constantly injected and/or re-accelerated in-situ with great efficiency over the cluster volume.

\begin{table}
\caption{Best fit parameters of the spectral energy distribution of the non-thermal emission 
from the Ophiuchus cluster. The reported uncertainties are at 1$\sigma$ level.}
\begin{center}
\begin{tabular} {ccc}
\hline
Parameter &     Value          & Description \\
\hline\\
$K_0$         & $0.53^{+6.16}_{-0.52}$ ~~cm$^{-3}$ & Energy spectrum normalization\\

$p$           & $3.83^{+0.38}_{-0.44}$        & Energy spectrum index \\

$\gamma_{min}$ & $\lesssim 700$; fixed to 300         & Low-energy cut off\\

$\gamma_{max}$ &  $3.83_{0.66}^{+3.16} \times 10^4$        & High-energy cut off\\

$B_{V}$       & $0.31^{+0.11}_{-0.07}$  ~~$\mu G$ & Volume-averaged magnetic field\\

$\chi^2$/d.o.f.  & 0.36/4               &\\

\hline
\end{tabular}
\label{tab4}
\end{center}
\end{table}

It is important to stress that the interpretation of the SED presented above relies on a simplified cluster 
model in which the
relevant physical properties, $N(\gamma,\theta)$ and $B$, represent volume-averaged quantities. It is clear, 
however, that
a much detailed modelling is needed to explain the spatial variations of the spectral index
observed across the mini-halo. In particular, the systematic increase of the mini-halo spectral index  
 with radius shown in the bottom panel of Fig.\ref{fig7} may indicate that
 the high-frequency break $\nu_{max}$, and hence either $\gamma_{max}$ and/or
the magnetic field $B$ (see Eq.\,\ref{nusyn}), decrease from the cluster centre outward.
Another possibility is that just the slope, $p$, and the normalization, $K_{0}$, of the energy spectrum 
of the non-thermal electrons are changing with radius. This could be supported by the radio spectrum of patch E, which does not
 show evidence for a strong spectral curvature (see Fig.\,\ref{fig7}). 
Disentangling between these scenarios is not easy in the view of the current data. Future, spatially resolved, analyses of both the 
radio and HXR emissions could shed light on this important issue.
 
\begin{figure*}
\includegraphics[width=18 cm]{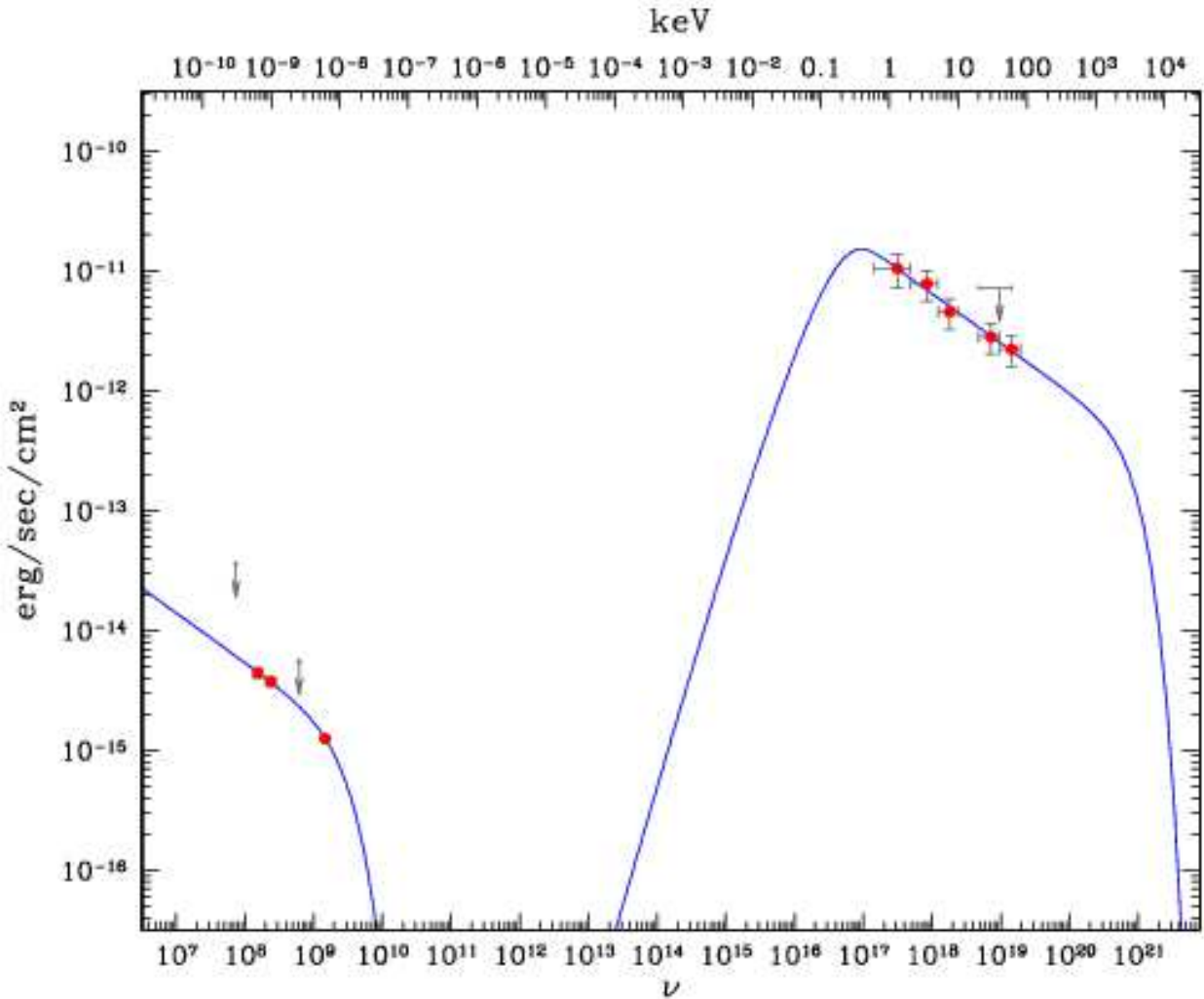}
\caption{Spectral energy distribution of the non-thermal emission  
(plotted as $\nu\cdot S_{\nu}$ versus $\nu$) from the Ophiuchus cluster, using data from GMRT (153 and 240 MHz) and VLA (1477 MHz) 
from Table 3, and data from \xmm\ and \intgr\ (see text) in the X-ray range.
Upper limits at 74 MHz (VLSS), 614 MHz (this work), and 
in the 20-60 keV band (Swift, Ajello et al. 2009) are also displayed.}
\label{sed}
\end{figure*}

\section{Conclusions}
In a search for diffuse radio emission in relaxed, cool-core, galaxy clusters at 1.4 GHz, Govoni
 et al. (2009) found the presence of a mini-halo surrounding the faint central 
point-like radio source in the Ophiuchus cluster of galaxies. Murgia et al. (2009) analyzed the
radio properties of this diffuse radio source in comparison to other mini-halos and radio halos
 known in the literature and found that Ophiuchus is characterized by brightness and size
much similar to that of the smaller halos rather than to that of the prototypical mini-halo
 in the Perseus cluster (e.g. Burns et al. 1992). In this work we presented a study of the radio emission of the Ophiuchus cluster of
galaxies at low radio frequencies performed at 153, 240, and 614 MHz with the GMRT.

The mini-halo is clearly detected at 153 and 240 MHz, the frequencies at
which we reached the best sensitivity to the low-surface brightness
diffuse emission, while at 614 MHz we only derived an upper limit to the mini-halo emission. 
By combining these images with the VLA data at 1477 MHz from Govoni et
 al. (2009) and with the VLSS upper limit at 74 MHz, we derived the spectral index of the mini-halo. 

Globally, the mini-halo has a low-frequency spectral index of 
$\alpha_{240}^{153}\simeq 1.4\pm0.3$, with a hint of steepening at higher frequencies.
Moreover, we found indications that the high-frequency spectral index  $\alpha_{1477}^{240}$ increases 
with the increasing distance from the cluster centre.
The most prominent feature at low frequencies is a patch of 
diffuse steep spectrum emission located at 
about 5\arcmin~south-east from the cluster centre.

The observed radio spectral index is in agreement with that obtained by 
modelling the non-thermal hard X-ray emission in this
cluster of galaxies. We assume that the X-ray component arises from inverse-Compton 
scattering between the photons of the cosmic microwave background and a population of non-thermal electrons 
which are isotropically distributed and whose energy spectrum is a power law with index $p$. We derive that
the electrons energy spectrum should extend from a minimum Lorentz factor of $\gamma_{min}\lesssim 700$ up to 
a maximum Lorentz factor of $\gamma_{max}\simeq 3.8 \times 10^{4}$ with an index $p=3.8\pm 0.4$ and that the 
volume-averaged strength for a completely disordered intra-cluster magnetic field is $B_V\simeq 0.3\pm 0.1\, \mu$G.
Given such a magnetic field strength, only high-energy electrons with $\gamma> 10^4$ can emit in the observed radio frequency window 
while the HXR emission would be tracing relativistic electrons 
of lower energy, with characteristic Lorentz factors in the range $10^3$ $-$ $10^4$. Indeed, the 
radio and HXR emissions are not tracing exactly the same particles. Nevertheless, the low-frequency radio spectral 
index measured in this work is consistent with the scenario in which
the energy spectrum of the synchrotron electrons most probably belongs to the extrapolation at higher energies
of the power law energy distribution of the electrons radiating in the HXR band through the inverse Compton process.

In addition to the mini-halo spectrum, we also analyzed the properties of the cluster discrete sources.
 Specifically, source A, the radio source associated to the central cD galaxy, source B, the brightest radio source 
in the field located at about 10\arcmin~north to the cluster centre, and the two NATs C and D located in the cluster outskirts.
Source A is point-like at our highest resolution of 7\arcsec~ (corresponding to about 3.8 kpc) 
and its spectral index is $\alpha\simeq 0.6$, a typical value for radio sources. 
Source B is an extended source with no obvious optical identification and a 
rather peculiar morphology which makes its classification very uncertain.
The global spectral index of the source is $\alpha\simeq 1.01\pm 0.05$. This is a 
quite usual value for active radio galaxies, which makes the interpretation of this object even 
more puzzling. In fact, although the distorted morphology of this radio 
source recall that of extreme relic sources in clusters of galaxies (see 
e.g. Slee et al. 2001), on the basis of its radio spectrum it cannot be 
classified as an ultra-steep spectrum source. A possibility could be that 
source B is a relic radio source revived by the adiabatic compression caused by a 
shock wave or a bulk gas motion propagating thought the ICM (En{\ss}lin \& Gopal-Krishna 2001). 
However, the relation of this peculiar radio source with the mini-halo remains, at the moment, 
unclear.
Finally, the analysis of the radio morphology and spectral properties of the tailed sources C and D 
confirm that their host galaxies are moving at 
high velocity with respect to the ICM either because they are infalling 
individually towards the cluster centre or because they are part of 
merging sub-clusters.

\begin{acknowledgements} 
We acknowledge the anonymous referee for helpful comments that improved the paper.
We thank the staff of the GMRT who have made these observations possible. GMRT is run by the National 
Centre for Radio Astrophysics of the Tata Institute of Fundamental Research.
This work is part of the ``Cybersar''  Project, which is managed by the COSMOLAB Regional Consortium with
the financial support of the Italian Ministry of University and Research (MUR), in the context 
of the ``Piano Operativo Nazionale Ricerca Scientifica, Sviluppo Tecnologico, Alta Formazione (PON  2000-2006)''.
This research was partially supported by ASI-INAF I/088/06/0 -High Energy 
Astrophysics and PRIN-INAF 2008. CF acknowledges financial support by the Agence Nationale de la 
Recherche through grant ANR-09-JCJC-0001-01. JN is supported by the Academy of Finland. The National Radio 
Astronomy Observatory (NRAO) is a facility of the National Science 
Foundation, operated under cooperative agreement by Associated 
Universities, Inc.
\end{acknowledgements}


\begin{thebibliography}{}

\bibitem{} Ajello, M., Rebusco, P. Cappelluti, N., et al., 2009, ApJ, 690, 367

\bibitem{} Arnaud, M., B\"ohringer, H., Jones, C., et al., 2009, arXiv:0902.4890

\bibitem{} Bird, A. J. , Barlow, E.J., Bassani, L., et al., 2006, ApJ, 636, 765

\bibitem{} Bliton, M., Rizza, E., Burns, J.~O., et al., 1998, \mnras, 301, 609 

\bibitem{} Blumenthal, G.~R., \& Gould, R.~J.\ 1970, Reviews of Modern Physics, 42, 237 

\bibitem{} Boselli, A., \& Gavazzi, G., 2006, PASP, 118, 517

\bibitem{} Burns, J.O., Sulkanen, M.E., Gisler, G.R., \& Perley, R.A. 1992, \apj, 388, L49

\bibitem{} Burns, J.~O., Hallman, E.~J., Gantner, et al. \ 2008, \apj, 675, 1125

\bibitem{} Cohen, A.~S., Lane, W.~M., Cotton, W.~D., et al., 2007, \aj, 134, 1245 

\bibitem{} Colafrancesco, S., \& Marchegiani, P.\ 2009, \aap, 502, 711 

\bibitem[]{} Djorgovski, S., Thompson, D.~J., de Carvalho, R.~R., \& Mould, J.~R.\ 1990, \aj, 100, 599 

\bibitem{} Dursi, L.J, \& Pfrommer, C., 2008, ApJ, 677, 993


\bibitem{} Eckert, D., Produit, N., Paltani, S., Neronov, A., \& Courvoisier T.J.-L., 2008, A\&A, 479, 27

\bibitem{} En{\ss}lin, T.~A., \& Gopal-Krishna 2001, \aap, 366, 26 

\bibitem{} Ferrari, C., Govoni, F., Schindler, S., Bykov, A. M., \&  Rephaeli, Y., 2008. SSRv, 134, 93

\bibitem{} Fujita, Y., Hayashida, K., Nagai, M., et al., 2008, \pasj, 60, 1133 

\bibitem{} Fusco-Femiano, R., dal Fiume, D., Orlandini, et al. \ 2003, Astronomical Society of the Pacific Conference Series, 301, 109 

\bibitem{} Gitti, M., Brunetti, G., Feretti, L., \& Setti, G., 2004,  A\&A, 417, 1

\bibitem{} Gitti, M., Ferrari, C., Domainko, W., Feretti, L., \& Schindler, S., 2007, A\&A, 470, 25

\bibitem{} Giovannini, G., Bonafede, A., Feretti, et al., \ 2009, arXiv:0909.0911 

\bibitem{} Govoni, F., Murgia, M., Markevitch, M., et al., 2009, A\&A, 499, 371

\bibitem[]{} Jaffe, W.~J., \& Perola, G.~C.\ 1973, \aap, 26, 423 

\bibitem{} Johnston, M.D., Bradt, H.V., Doxsey, R.E., et al., 1981, ApJ, 245, 799

\bibitem{} Kawano, N., Fukazawa, Y., Nishino, S., et al.\ 2009, \pasj, 61, 377 

\bibitem{} Kravtsov, A., Gonzalez, A., Vikhlinin, A., et al., 2009, arXiv:0903.0388

\bibitem{} Markevitch, M., \& Vikhlinin, A., 2007, Physics Reports, Volume 443, Issue 1, p. 1-53

\bibitem{} Melrose, D.~B.\ 1968, \apss, 2, 171 

\bibitem{} Million, E.T., Allen, S.W., Werner, N., \& Taylor, G.B., 2009, arXiv:0910.0025

\bibitem{} Murgia, M., Govoni, F., Markevitch, M., et al., \ 2009, \aap, 499, 679 

\bibitem{} Nevalainen, J., Oosterbroek, T., Bonamente, M., \& Colafrancesco, S.\ 2004, \apj, 608, 166 

\bibitem{} Nevalainen, J., Eckert, D., Kaastra, J., Bonamente, M., \& Kettula, K., 2009, arXiv:0910.1364

\bibitem[]{} Pacholczyk, A.~G.\ 1970, Series of Books in Astronomy and Astrophysics, San Francisco: Freeman, 1970, 

\bibitem[]{} Parma, P., Murgia, M., Morganti, et al. \ 1999, \aap, 344, 7 

\bibitem{} Parrish, I.J., Quataert, E., \& Sharma, P., 2009, ApJ, 703, 96

\bibitem{} Petrosian, V., Bykov, A., \& Rephaeli, Y.\ 2008, Space Science Reviews, 134, 191 

\bibitem{} P{\'e}rez-Torres, M.~A., Zandanel, F., Guerrero, M.~A., et al., 2009, \mnras, 396, 2237 

\bibitem{} Profumo, S.\ 2008, \prd, 77, 103510 

\bibitem{} Rephaeli, Y., Nevalainen, J., Ohashi, T., \& Bykov, A.~M.\ 2008, Space Science Reviews, 134, 71 

\bibitem{} Rybicki, G.~B., \& Lightman, A.~P.\ 1979, New York, Wiley-Interscience, 1979. 393 p.

\bibitem{} Sarazin, C.~L.\ 1999, \apj, 520, 529 

\bibitem{} Wakamatsu, K., Malkan, M.A., Nishida, M.T., et al., 2005, Nearby Large-Scale Structures and the Zone of Avoidance, 329, 189 


\end{thebibliography}
\end{document}